\documentclass[runningheads]{llncs}
\usepackage{graphicx}
\usepackage[T1]{fontenc}    
\usepackage{afterpage}      
\usepackage{csquotes}
\usepackage{float}
\usepackage{url}
\floatstyle{ruled}
\newfloat{program}{t}{lop}
\floatname{program}{Program}
\usepackage{fancyvrb}
\usepackage{parcolumns}
\usepackage{wrapfig}
\usepackage{lipsum}
\usepackage{hyphenat}
\usepackage{alltt,amsmath,amssymb}
\usepackage{cite}

\usepackage{color}

\newcommand{\pamper}{\texttt{PaMpeR}}

\newcommand{\etal}{\textit{et al.}}

\newcommand{\auto}{\texttt{auto}}
\newcommand{\simp}{\texttt{simp}}
\newcommand{\induct}{\texttt{induct}}
\newcommand{\introclasses}{\texttt{intro\_classes}}
\newcommand{\introlocales}{\texttt{intro\_locales}}
\begin{document}
%
\title{Simple Dataset for Proof Method Recommendation in Isabelle/HOL
\thanks{
This work was supported by the European Regional Development Fund under the project 
AI \& Reasoning (reg. no.CZ.02.1.01/0.0/0.0/15\_003/0000466)
and by NII under NII-Internship Program 2019-2nd call.}
}
%
%
\author{Yutaka Nagashima\inst{1}\inst{2}
\orcidID{0000-0001-6693-5325}}
\authorrunning{Y. Nagashima}
%
\institute{
Czech Technical University in Prague,
Prague, Czech Republic
\email{Yutaka.Nagashima@cvut.cz}
\and
University of Innsbruck, Innsbruck, Austria}

\maketitle              
\begin{abstract}
Recently, a growing number of researchers have 
applied machine learning to assist users of interactive theorem provers.
However, the expressive nature of underlying logics and esoteric structures of proof documents impede machine learning practitioners,
who often do not have much expertise in formal logic, let alone Isabelle/HOL,
from achieving a large scale success in this field. 
In this data description, we present a simple dataset that 
contains data on over 400k proof method applications along with over 100 extracted features
for each in a format 
that can be processed easily without any knowledge about formal logic.
Our simple data format allows machine learning practitioners to
try machine learning tools to predict proof methods in Isabelle/HOL
without requiring domain expertise in logic.
\end{abstract}

\section{Introduction}
As our society relies heavily on software systems,
it has become essential to ensure that 
our software systems are trustworthy.
Interactive theorem provers (ITPs), such as Isabelle/HOL \cite{isabelle}, 
allow users to specify desirable functionalities of a system and 
prove that the corresponding implementation is correct in terms of the specification.

A crucial step in developing proof documents in ITPs is 
to choose the right tool for a proof goal at hand.
Isabelle/HOL, for example, comes with more than 100 proof methods.
Proof methods are sub-tools inside Isabelle/HOL.
Some of these are general purpose methods, such as \auto{} and \simp{}.
Others are special purpose methods, such as \introclasses{} and \introlocales{}.
The Isabelle community provides various documentations \cite{isabelle} and on-line supports
to help new Isabelle users learn when to use which proof methods.

Previously, we developed \pamper{} \cite{pamper}, 
a \underline{p}roof \underline{m}ethod \underline{r}ecommendation tool for Isabelle/HOL.
Given a proof goal specified in a proof context,
\pamper{} recommends a list of proof methods likely to be suitable for the goal.
\pamper{} learns which proof method to recommend to what kind of proof goal
from proof documents in Isabelle's standard library and the Archive of Formal Proofs \cite{AFP}.

The key component of \pamper{} is its elaborate feature extractor.
Instead of applying machine learning algorithms to Isabelle's proof documents directly,
\pamper{} first applies 113 assertions to 
the pair of a proof goal and its underlying context.
Each assertion checks a certain property about the pair and returns a boolean value.
Some assertions check 
if a proof goal involves certain constants or types defined in the standard library.
Others check the meta-data of constants and types appearing in a goal.
For example, one assertion checks
if the goal has a term of a type defined with the \texttt{codatatype} keyword.

When developing \pamper{}, 
we applied these 113 assertions to the proof method invocations appearing in the proof documents
and constructed a dataset consisting of 425,334 unique data points.

Note that this number is strictly smaller than all the available proof method
invocations in Isabelle2020 and the Archive of Formal Proofs in May 2020,
from which we can find more than 900k proof method invocations.
One obvious reason for this gap is the ever growing size of the available proof documents.
The other reason is that we are intentionally ignoring compound proof methods
while producing data points.
We decided to ignore them
because they may pollute the database by introducing
proof method invocations that are eventually backtracked by Isabelle.
Such backtracking compound methods may reduce the size of proof documents
at the cost of introducing backtracked proof steps, 
which are not necessary to complete proofs.
Since we are trying to recommend proof methods appropriate to complete a proof search,
we should not include data points produced by such backtracked steps.

We trained \pamper{} by constructing regression trees \cite{decision_tree} from this dataset.
Even though our tree construction is based on a fixed height
and we did not take advantage of modern development of machine learning research,
our cross evaluation showed
\pamper{} can correctly predict experts' choice
of proof methods for many cases.
However, decision tree construction based on a fixed height is an old technique
that tends to cause overfitting and underfitting.
We expect that one can achieve better performance by applying
other algorithms to this dataset.

In the following we present the simple dataset we used to train \pamper{}.
Our aim is to provide a dataset that is publicly available at Zenodo \cite{zenodo}
and easily usable 
for machine learning practitioners
without backgrounds in theorem proving,
so that they can exploit the latest development of machine learning
research without being hampered by technicalities of theorem proving.

\section{The \pamper{} Dataset}

Each data point in the dataset consists of the following three entries:
\begin{itemize}
    \item the location of a proof method invocation,
    \item the name of the proof method used there,
    \item an array of \verb|0|s and \verb|1|s expressing the proof goal and its context.
\end{itemize}
\noindent
The following is an example data point:
\begin{verbatim}
Functors.thy119 simp 1,0,0,0,0,0,0,0,0,0,0,0,0,0,1,...
\end{verbatim}
\noindent
This data point describes that in the theory file named \texttt{Functors.thy},
a proof author applied the \simp{} method in line 119 to a proof goal represented
by the sequence of \texttt{1}s and \texttt{0}s where
\texttt{1} indicates the corresponding assertion returns true while
\texttt{0} indicates the otherwise.

This dataset has important characteristics worth mentioning.
Firstly, this dataset is heavily imbalanced in terms of occurrences of proof methods.
Some general purpose methods, such as \auto{} and \simp{}, appear far more
often than other lesser known methods:
each of \auto{} and \simp{} accounts more than 25\% of all proof method invocations
in the dataset, whereas
no proof methods account for more than 1\% of invocations
except for the 15 most popular methods.

Secondly, this dataset only serves to learn what proof methods to apply,
but it does not describe how to apply a proof method.
None of our 113 assertions examines arguments passed to proof methods.
For some proof methods, notably the \induct{} method, 
the choice of arguments is the hardest problem to tackle,
whereas some methods rarely take arguments at all.
We hope that 
users can learn what arguments to pass to proof methods from 
the use case of these methods in existing proof documents
once they learn which methods to apply to their goal.

Thirdly, it is certainly possible that 
\pamper{}'s feature extractor misses out certain information essential
to accurately recommend some methods.
This dataset was not built to preserve the information in the original proof documents:
we built the dataset, so that we can effectively apply machine learning algorithms to
produce recommendations.

Finally, this dataset shows only one way to prove a given goal,
ignoring alternative possible approaches to prove the same goal.
Consider the following goal: \texttt{"True $\lor$ False"}.
Both \auto{} or \simp{} can prove this goal equally well;
however, 
if this goal appeared in our dataset
our dataset would show only the choice of the proof author, say \auto{},
ignoring alternative proofs, say \simp{}.

One might guess that we could build a larger dataset that also includes
alternative proofs by trying to complete a proof using various methods,
thus converting this problem into a multi-label problem.
That approach would suffer from two problems.
Firstly, there are infinitely many ways to apply methods
since we often have to apply multiple proof methods in a sequence to prove a conjecture.
Secondly,
some combinations of methods are not appropriate 
even though they can finish a proof in Isabelle.
For example, the following is an alternative proof for the aforementioned proposition:
\begin{alltt}
lemma "True \(\lor\) False" apply(rule disjI1) apply auto done
\end{alltt}
\noindent
This is a valid proof script, with which Isabelle can check the correctness of the conjecture; however,
the application of the \texttt{rule} method is hardly appropriate
since the subsequent application of the \auto{} method can discharge 
the proof without the preceding \texttt{rule}.
For these reasons we take the proof methods chosen by human proof authors as 
the correct choice while ignoring other possibilities.

\section{Overview of 113 Assertions}

The 113 assertions we used to build the dataset roughly fall into the following two categories:
\begin{itemize}
    \item[1.] assertions that check terms and types appearing in the first sub-goal, and
    \item[2.] assertions that check how such terms and types are defined in the underlying proof context.
\end{itemize}

The first kind of assertions directly check the presence of constructs defined in the standard library.
For example, the 56th assertion checks if the first sub-goal contains 
\texttt{Filter.eventually}, which is a constant defined in the standard library
since the presence of this constant may be a good indicator to recommend
the special purpose proof method called \texttt{eventually\_elim}.
A possible limitation of these assertions is that
these assertions cannot directly check the presence of user-defined constructs
because such constructs may not even exist when we develop the feature extractor.

The second kind of assertions address this issue by checking how 
constructs appearing in the first sub-goal are defined in the proof context.
For example, the 13th assertion checks if the first sub-goal involves a constant that 
has one of the following related rules: 
the \texttt{code} rule, the \texttt{ctr} rule, and the \texttt{sel} rule.

These related rules are derived by Isabelle when human engineers define new constants
using the \texttt{primcorec} keyword,
which is used to define primitively corecursive functions.
Since this assertion checks how constants are defined in the background context,
it can tell that the proof goal at hand is a coinductive problem.
Therefore, if this assertion returns \texttt{true},
maybe the special purpose method called \texttt{coinduct} would be useful,
since it is developed for coinductive problems.
The advantage of this assertions is that
it can guess if a problem is a coinductive problem or not,
even though we did not have that problem at hand
when developing the assertion.

Due to the page limit, we expound the further details of 
the 113 assertions in our accompanying Appendix \cite{appendix}.

\section{The Task for Machine Learning Algorithms}

The task for machine learning algorithms is to predict the name 
of a promising proof method from the corresponding array of boolean values.
Since we often have multiple equivalently suitable methods for a given proof goal,
this learning task should be seen as a multi-output problem:
given an array of boolean values machine learning algorithms should 
return multiple candidate proof methods rather than only one method.
Furthermore, this problem should be treated as a regression problem rather 
than a classification problem, so that
users can see numerical estimates about how likely each method is suitable
for a given goal.

\section{Conclusion and Related Work}
We presented our dataset for proof method recommendation in Isabelle/HOL.
Its simple data format allows machine learning practitioners to try out various algorithms
to improve the performance of proof method recommendation.

Kaliszyk \etal{} presented HolStep \cite{hostelp}, a dataset based on proofs for HOL Light  \cite{hollight}.
They developed the dataset from  a multivariate analysis library \cite{euclidean} 
and the proof of the Kepler conjecture \cite{kepler}.
They built HolStep for for various tasks, which does not include proof method prediction.
While their dataset explicitly describes the text representations of conjectures 
and dependencies of theorems and constants,
our dataset presents only the essential information about proof documents
as an array of boolean values.

Blanchette \etal{} mined the Archive of Formal Proofs \cite{mining_afp}
and investigated the nature of proof developments, such as the size and complexity of proofs \cite{mining_afp_dan}.
Matichuk \etal{} also studied the Archive of Formal Proofs 
to understand leading indicators of proof size
\cite{mining_afp_dan}.
Neither of their projects aimed at suggesting how to write proof documents:
to the best of our knowledge we are the first to mine a large repository of ITP proofs
using hand crafted feature extractors.

Our dataset does not contain information useful to predict 
what arguments to pass to each method.
Previously we developed, \texttt{smart\_induct} \cite{smart_induct}, to address 
this problem for the \induct{} method in Isabelle/HOL,
using a domain-specific language for logical feature extraction \cite{lifter}.

Recently a number of researchers have developed meta-tools that 
exploit existing proof methods and tactics and brought stronger proof automation to ITPs
\cite{tactictoe, psl, pgt, holist, ml4pg2, sepia}.
We hope that our dataset helps them improve the performance of such meta-tools for Isabelle/HOL.

%


%
%
 \bibliographystyle{splncs04}
 \bibliography{bibfile}
%

\end{document}